# Computational ghost imaging using a field-programmable gate array

**Ikuo Hoshi,** *  **Tomoyoshi Shimobaba, Takashi Kakue, and Tomoyoshi Ito**

*1Graduate School of Engineering, Chiba University, 1-33, Yayoi-cho, Inage-ku, Chiba, Japan*
*\* aeka2345@chiba-u.jp*

**Abstract:** Computational ghost imaging is a promising technique for single-pixel imaging because it is robust to disturbance and can be operated over broad wavelength bands, unlike common cameras. However, one disadvantage of this method is that it has a long calculation time for image reconstruction. In this paper, we have designed a dedicated calculation circuit that accelerated the process of computational ghost imaging. We implemented this circuit by using a field-programmable gate array, which reduced the calculation time for the circuit compared to a CPU. The dedicated circuit reconstructs images at a frame rate of 300 Hz.

## 1. Introduction

Ghost imaging (GI) is an imaging method that has been intensively studied in recent years [1-4]. Unlike the usual imaging that uses charge-coupled devices, GI uses a single-pixel device as the light-receiving element. In GI, the object is illuminated by using light having spatially random patterns; then, the light that passes through the objects (or the light reflected by the objects) is corrected by the lens. These lights are called object lights. The intensity of the object light is detected by a single-pixel element. Finally, the object image is reconstructed by calculating the correlation between the obtained object light intensities and the random illumination patterns used to obtain the object light intensities. Researchers have proposed GI-based methods for calculating the light-intensity distribution of the random illumination patterns on a computer [5, 6]; this is called computational GI.

Computational GI is advantageous for measurements over broad wavelength bands; it is robust to disturbance and simplifies the optical system. These characteristics are expected to be applied in a wide range of fields such as bio imaging [7], remote sensing [8], and encryption [9]; computational GI is also helpful in taking three-dimensional measurements [10]. However, this method also has its disadvantages—the image quality of the reconstructed image is poor; the measurement time is high; and the reconstruction calculation is time consuming. Research has been conducted for improving the image quality by using modified correlation calculation [11, 12], compressive sensing [13], and deep learning [14, 15]. Research has also been conducted for shortening the measurement time [7, 16, 17].

To accelerate the reconstruction calculation, we have designed a calculation circuit for computational GI, which can calculate the pixels of the reconstructed image in parallel. We implemented this circuit in a field-programmable gate array (FPGA). The object light intensities obtained from the optical system were input to the FPGA, and a reconstructed image was obtained by calculating the correlation. The reconstruction time in FPGA for images having $32 \times 32$ pixels was 3 ms. This implies that this circuit can reconstruct images at a frame rate of 300 Hz or more.

In Section 2, we describe the principle of computational GI, and we describe the calculation circuit used for the computational GI. In Section 3, we show the results obtained by implementing the proposed circuit into the FPGA. We compare the calculation speed and evaluate the image quality of the reconstructed images. In Section 4, we summarize this research.

## 2. Hardware implementation of computational ghost imaging

Figure 1 presents a schematic of the computational GI used in this research. By using a digital mirror device (DMD) projector to illuminate an object with random illumination patterns, we obtained the time-series data of the object light intensities by using a photo detector and an analog-to-digital (AD) converter. The time-series data were sent to the memory of the FPGA. The parallel processing of the reconstruction calculation on the FPGA improved the speed. A universal serial bus (USB) interface was used for the communication between a personal computer and the FPGA board.

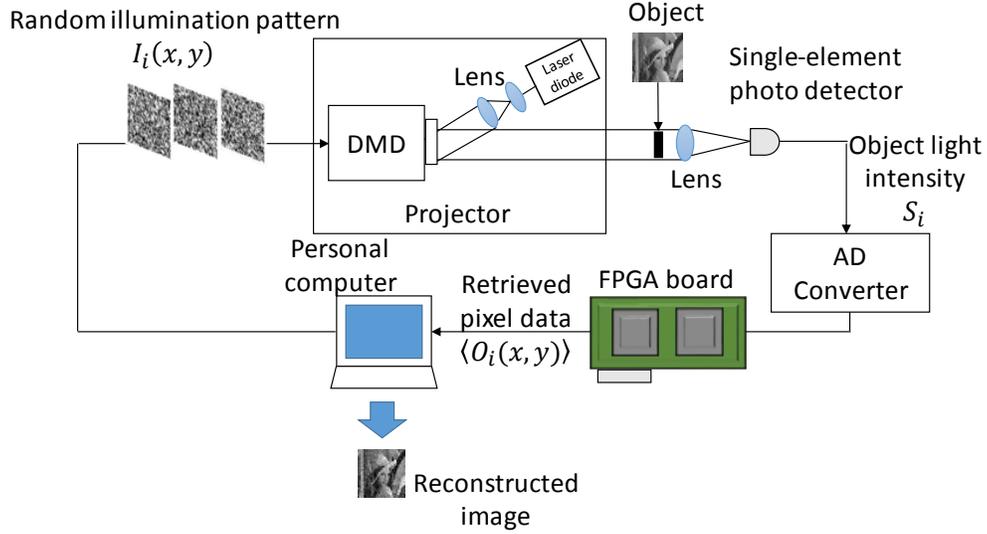

Fig. 1. Optical system with the FPGA for the computational ghost imaging.

After the random illumination pattern was passed through the object, the object light intensity was collected by the lens and detected by the photo detector. The detected object's light intensity $S_i$ is given as

$$S_i = \iint I_i(x,y)T(x,y)dxdy, \quad (1)$$

where $I_i(x,y)$ is the distribution of the random illumination pattern, and $T(x,y)$ is the transmittance of the object. The intensity of the random illumination pattern $R_i$ is given as

$$R_i = \iint I_i(x,y)dxdy. \quad (2)$$

The following formula called the differential GI (DGI) [11, 12] of a computational GI was used for the reconstruction:

$$\langle O_i(x,y) \rangle = \langle S_i I_i(x,y) \rangle - \frac{\langle S_i \rangle}{\langle R_i \rangle} \langle R_i I_i(x,y) \rangle, \quad (3)$$

where $\langle O_i(x,y) \rangle$ represents the reconstructed image, and $\langle \cdots \rangle$ represents the ensemble average. In Eq. (3), $\langle S_i I_i(x,y) \rangle$ and $\langle R_i I_i(x,y) \rangle$ require $n \times x \times y$ times calculations, where $n$ is the number of the random illumination patterns, and $x \times y$ is the number of the pixels. These operations are the most time consuming in DGI. $\langle R_i \rangle$ and $\langle R_i I_i(x,y) \rangle$ do not depend on objects;

therefore, they can be calculated in advance. Instead of using central processing units (CPUs), we designed a dedicated circuit to accelerate the computation of Eq. (3).

We compared the image quality obtained by using the original computational GI [5] and the DGI under the same conditions. The reconstructed images are shown in Fig. 2. Figure 2(a) is the image reconstructed by using the computational GI, and Fig. 2(b) is the image reconstructed by using the DGI. We adopted the DGI for the hardware implementation because the image quality of the DGI was obviously better than that of the computational GI.

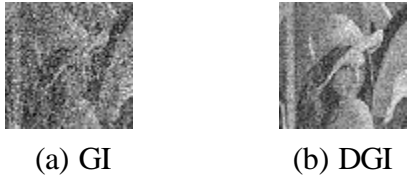

(a) GI    (b) DGI

Fig. 2. Comparison of the images obtained by using (a) computational GI and (b) DGI.

In Eq. (3), the division by $\langle R_i \rangle$ is a bottleneck in the hardware implementation. To simplify the hardware implementation, we reformulate Eq. (3) as follows:

$$\langle R_i \rangle \langle O_i(x,y) \rangle = \langle R_i \rangle \cdot \langle S_i I_i(x,y) \rangle - \langle S_i \rangle \cdot \langle R_i I_i(x,y) \rangle. \quad (4)$$

The dedicated circuit generates random illumination patterns that are the same as the patterns displayed on the DMD projector. The pseudo-random number generators—linear congruential generators (LCGs), Mersenne Twister (MT), and the maximum length sequence (hereinafter "M-sequence")—generate random illumination patterns. The reconstructed images generated by each method are shown in Fig. 3. Figures 3(a), 3(b), and 3(c) are the reconstructed images obtained by using LCGs, MT, and M-sequence, respectively. There were almost no differences in the image quality. In terms of the hardware implementation, we selected M-sequence as the pseudo-random number generator.

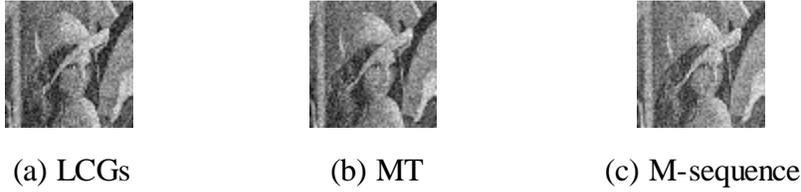

(a) LCGs    (b) MT    (c) M-sequence

Fig. 3. Comparison of the reconstructed images using LCGs, MT, and the M-sequence.

### 3. Designing the calculation circuit

The schematic of the dedicated circuit is shown in Fig. 4. This circuit has three parts: a receiver unit, a calculation unit, and a transmitter unit. The receiver unit and the transmitter unit are the USB transmission circuits between the host computer and the FPGA. The calculation unit reconstructs images with 32 × 32 pixels. We used Xilinx Artix-7 XC7A100T-2 as the FPGA. The dedicated circuit was operated at 100 MHz. The input data was the object light intensity obtained by the AD converter. The output data was the reconstructed image.

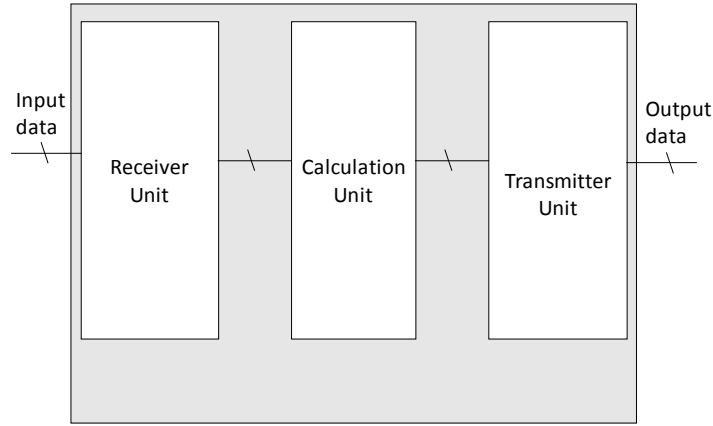

Fig. 4. Top configuration of the dedicated circuit.

The schematic of the calculation unit is shown in Fig. 5. All the arithmetic operations in the calculation circuit were developed using a fixed-point number. Figures 5 and 6 have several sets of three numbers in parentheses. Here, the first, second, and third numbers represent the sign bit, the number of bits of the integer part, and the decimal part of the fixed point number, respectively.

When the object light intensities were received, the calculation unit started by calculating the average $\langle S_i \rangle$. Then, the average $\langle S_i I_i(x,y) \rangle$ was calculated by the parallel calculator from the object light intensity $S_i$ saved in memory and from the random illumination pattern $I_i(x,y)$ that was generated from the random number generator. The calculated $\langle S_i I_i(x,y) \rangle$ was saved in a random access memory (RAM). Subsequently, $\langle O_i(x,y) \rangle$ was calculated from $\langle S_i I_i(x,y) \rangle$, $\langle S_i \rangle$, $\langle R_i I_i(x,y) \rangle$, and $\langle R_i \rangle$. $\langle R_i I_i(x,y) \rangle$ and $\langle R_i \rangle$ were pre-calculated in the host computer and stored in a table and registers, respectively. Finally, the calculation unit sent $\langle O_i(x,y) \rangle$ to the transmitter unit. Note that the $\langle R_i \rangle$ factor that appears on the left side of Eq. (4) can be omitted because it is a constant.

The details of the parallel calculator unit are shown in Fig. 6. This unit can simultaneously calculate 64 pixels in the reconstructed image because 64 calculation modules were operated in parallel. Figure 7 shows the details of the calculation module. As shown in Fig. 8, the 64 calculation modules process two lines of the reconstructed image (32 × 32 pixels); subsequently, they process the next two lines. All the calculated values were saved in the RAM shown in Fig. 5 via the multiplexer of Fig. 6.

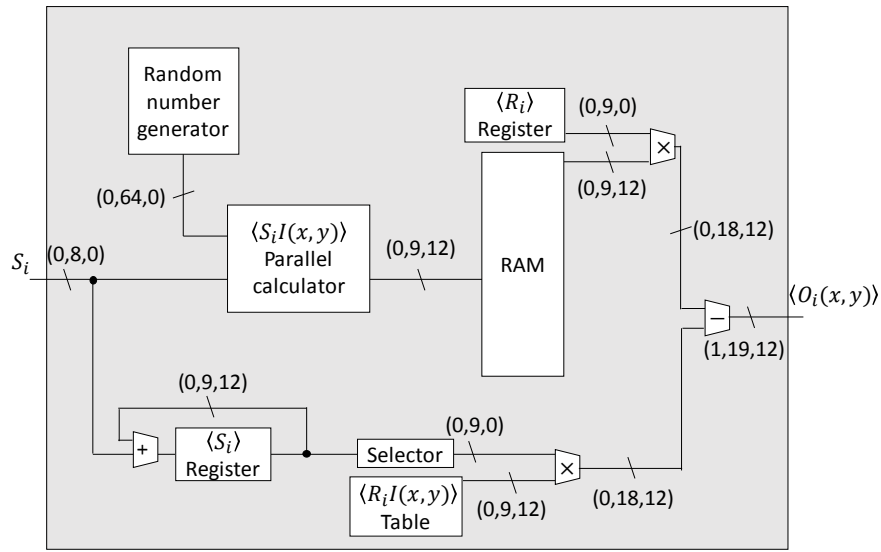

Fig. 5. Schematic of the calculation unit.

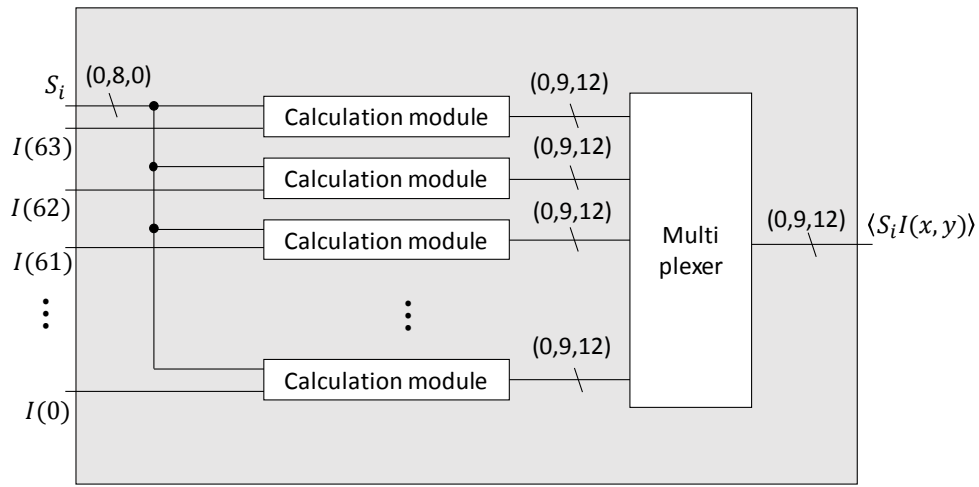

Fig. 6. Schematic of parallel calculator unit.

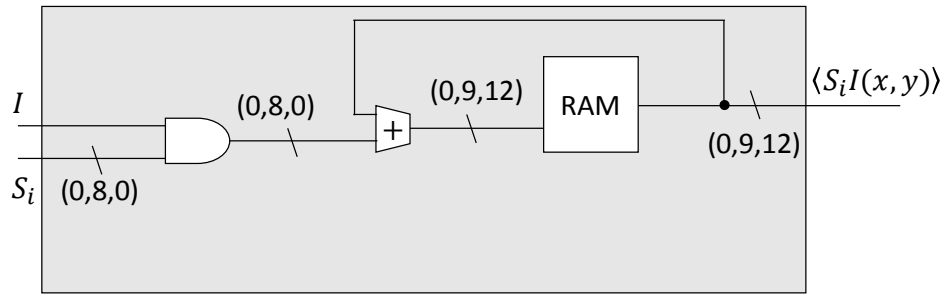

Fig. 7. Schematic of the calculation module.

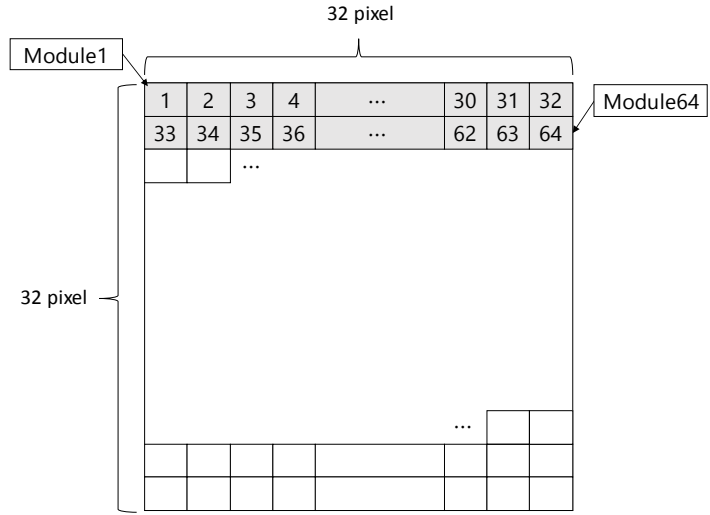

Fig. 8. Order of reconstruction calculation.

The random pattern generator using M-sequence is shown in Fig. 9. The boxes (called taps) with the notation M($s$) were implemented by flip-flops; $s$ indicates the index of the flip-flops. In this research, we generated a binary random number sequence by using the liner feedback shift register (LFSR). The feedback positions of the LFSR were determined by the longest polynomial [18]. The generator was necessary for producing 64-bit random numbers in parallel; therefore, the register needed to shift the current data to 64 bits per clock cycle.

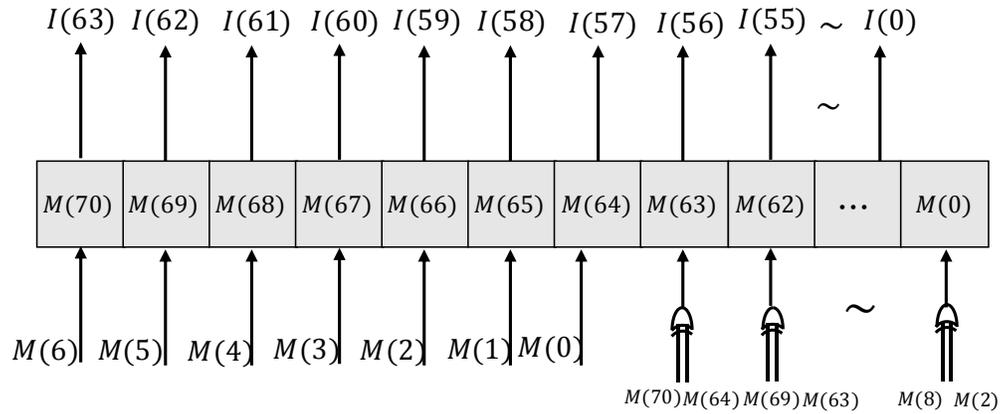

Fig. 9. M-sequence with 64-bit liner feedback shift register.

## 4. Result

In this study, the calculation time for image reconstruction when using a CPU was compared with the calculation time when using the FPGA. The number of random illumination patterns was 16,384. The calculation times of the FPGA were compared for 16 and 64 calculation modules. The transmission time between the FPGA and the host computer were not included in the calculation times. For the computing environment, we used Intel Core™ i5 4690 (clock frequency 3.50 GHz) as the CPU, a memory of 8.0 GB, Microsoft Windows 10 education as the operating system, and Microsoft Visual Studio C ++ 2015 as the compiler. The calculation times for the various devices are given in Table.1.

**Table 1. Calculation times.**

| Device | Calculation time [ms] |
|---|---|
| CPU | 45 |
| FPGA (16 calculation modules) | 10 |
| FPGA (64 calculation modules) | 3 |

From Table 1, it is clear that the calculation using the FPGA was faster than that using the CPU. In addition, as the number of parallel modules was increased, the calculation speed was improved. In the 64 calculation modules, the dedicated circuit could calculate the reconstructed image at 3 ms. In other words, the circuit reconstructed images at a frame rate of over 300 Hz. As the result, the parallelization was effective.

A few main advantages of using FPGAs are as follows:

(a) The object light intensity of the AD converter can be directly received by the FPGA without going through any CPU or operating systems;

(b) The reconstruction calculation can be performed without CPUs; and

(c) The power consumption is low.

In particular, the first advantage becomes very important in applications that require precise timing control, such as cytometry [7]. In such applications, it is necessary to accurately control the timing (latency) from the reception of the input signals to the image reconstruction. It is very difficult for CPUs and GPUs to control the latency because the calculation paths in CPUs and GPUs are complex; in addition, the paths are controlled by operating systems. However, FPGAs can accurately control the latency easily.

We evaluated the image quality of the reconstructed images obtained by the CPU and FPGA in the numerical simulations. The calculations in the FPGA used the fixed-point number. The calculations in the CPU used the floating-point number. Each reconstructed image is shown in Fig. 10. Figure 10(a) is the original image. Figure 10(b) is the reconstructed image obtained by the FPGA, and Fig. 10(c) is the reconstructed image obtained by the CPU. The peak signal-to-noise ratio (PSNR) and structural similarity (SSIM) were used for evaluating the image quality. The evaluated image quality is shown in Table 2. The qualitative and quantitative evaluations show that there is almost no difference between the image qualities.

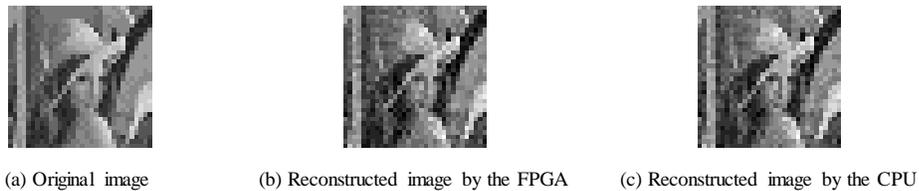

(a) Original image    (b) Reconstructed image by the FPGA    (c) Reconstructed image by the CPU

Fig. 10. Reconstructed images obtained by the CPU and FPGA.

**Table 2. Numerical evaluation of image quality.**

| Device | PSNR | SSIM |
|---|---|---|
| CPU (floating-point number) | 25.03 | 0.95 |
| FPGA (fixed-point number) | 23.62 | 0.94 |

Reconstructed images of the three objects using an actual optical system are shown in Fig. 10. We confirmed that the reconstructed images could be obtained by FPGA.

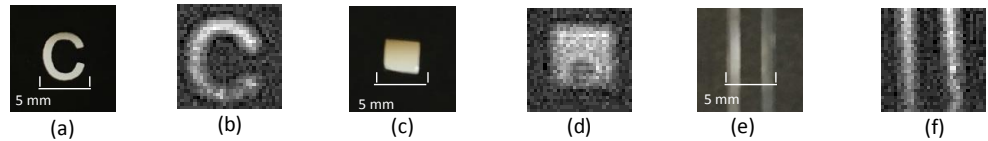

(a) (b) (c) (d) (e) (f)

Fig. 11. Original objects and reconstructed images using an actual optical system.

## 5. Conclusion

In this research, we designed a dedicated circuit to reduce the time taken for the image reconstruction by using computational GI. The dedicated circuit could reconstruct images at a frame rate of over 300 Hz. The image quality of the reconstructed images obtained by the FPGA was almost the same as that obtained by the CPU. We also confirmed that the FPGA could obtain reconstructed images in an actual optical system. The circuit scale of the FPGA used in this research was small. Larger reconstructed images could be obtained at higher speeds by using large-scale FPGAs. In this research, we used random pattern illumination. The image quality is expected to improve if the Fourier basis and Hadamard basis are used for illumination [16, 17]. In future, we plan to improve our dedicated circuit using upon the method.